\documentclass[twocolumn,pre,showpacs,floatfix,superscriptaddress]{revtex4}
\usepackage{graphicx}
\usepackage{amsmath}
\usepackage{amssymb}
\usepackage{color}

\begin{document}
\title{Percolation transitions with nonlocal constraint}

\author{Pyoung-Seop Shim}
\affiliation{Department of Physics, University of Seoul, Seoul 130-743, Korea}
\author{Hyun Keun Lee}
\affiliation{Department of Physics, University of Seoul, Seoul 130-743, Korea}
\author{Jae Dong Noh}
\affiliation{Department of Physics, University of Seoul, Seoul 130-743, Korea}
\affiliation{School of Physics, Korea Institute for Advanced Study,
Seoul 130-722, Korea}

\date{\today}

\begin{abstract}
We investigate percolation transitions in a nonlocal network model
numerically.
In this model, each node has an exclusive partner and a link is forbidden
between two nodes whose $r$-neighbors share any exclusive pair.
The $r$-neighbor of a node $x$ is defined as a set of
at most $N^r$ neighbors of $x$, where $N$ is the total number of nodes.
The parameter $r$ controls the strength of a nonlocal effect.
The system is found to undergo a percolation transition belonging
to the mean field universality class for $r< 1/2$.
On the other hand, for $r>1/2$, the system undergoes a peculiar
phase transition from a non-percolating phase to a quasi-critical phase
where the largest cluster size $G$ scales as $G \sim N^{\alpha}$ with 
$\alpha = 0.74~(1)$. In the marginal case with $r=1/2$, 
the model displays a percolation transition that does not belong to the 
mean field universality class.
\end{abstract}

\maketitle

A networked system is in a percolating phase when
a finite faction of nodes are interconnected via
links to form a percolating giant
cluster~\cite{Christensen,Stauffer94}. A percolating configuration requires
a macroscopic number of links, and the link density $l$ given by the ratio of
the number of links over the number of nodes is a control
parameter for a percolation transition. It is well known that there is a
threshold link density $l_c$ above which a percolating cluster appears.
Recently, the percolation in complex networks has been studied extensively 
for the understanding of emergent phenomena in complex
systems~\cite{Callaway00,Albert02}

A prototypical model for percolation transitions is the
random network model of Erd\H{o}s-R\'enyi~(ER) where each pair of nodes is
connected independently and uniformly with the same probability~\cite{ER}.
It displays a percolation transition at a critical
link density $l_c=1/2$, and its critical behavior exemplifies
the mean-field~(MF) universality class~\cite{Bray88,Stauffer94}.
The universality class is extended by generalizing the model in various
ways. For example, percolation transitions 
are studied in random networks with a
power-law degree distribution~\cite{Callaway00,Cohen00,Lee04}, in growing
networks~\cite{Callaway01,JKim02}, in correlated
networks~\cite{Noh07,Goltsev08,Agliari11}, and so on.

In most network models, a linking probability between nodes
depends only on a local information around
involved nodes. Recently, a nonlocal model was
proposed in Ref.~\cite{AchlioptasSCIENCE2009}, where
a node is linked with a probability depending on the whole cluster size
distribution.
This model exhibits an intriguing explosive percolation transition, and
a lot of works have been performed to clarify its 
nature~\cite{Ziff09,Cho10,Riodan11,Lee11,Manna11,Choi11,Tian12}.
These studies suggest that nonlocality may be 
an important ingredient for the universality class of percolation 
critical phenomena.

In this paper, we introduce a nonlocal network model under a so-called 
pair-exclusion~(PE) constraint and investigate its percolation transitions.
Each node is assigned to have its own exclusive counterpart. The
PE constraint means that a link between two nodes is forbidden
if two clusters containing each node share any exclusive pair.
Such a constraint was considered in evolving network models~\cite{SWKim09}
and mass aggregation models~\cite{SWKim10}. In the context of evolving 
networks, the PE corresponds to a constraint against 
self-loops~\cite{SWKim09}.
The PE type interaction may also be relevant in a social group formation 
process among individuals with mutual conflicts~\cite{Zachary77}. 
We apply the PE constraint to a percolation problem.

\begin{figure}[t]
\includegraphics*[width=\columnwidth]{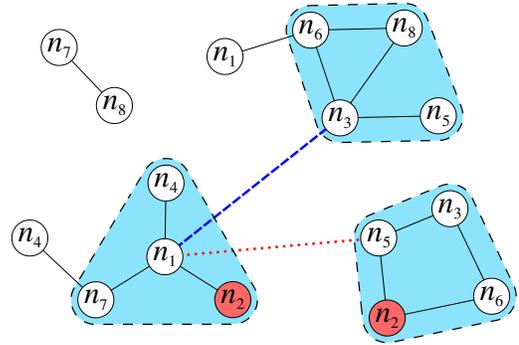}
\caption{(Color online) Illustration of dynamics of the model
with $N=16$ and $r=1/2$. Shaded regions represent the $r$-neighbors of nodes
of species $n_1$, $n_3,$ and $n_5$. One can add a link~(represented with a
dashed line) between nodes of
$n_1$ and $n_3$, while a link~(represented with a dotted line)
between nodes of $n_1$ and $n_5$ is forbidden
because of an exclusive pair of species $n_2$ represented with filled
circles.}\label{fig1}
\end{figure}

Consider a system consisting of $N$ nodes. In order to impose the PE
constraint, we assume that there are $N/2$ particle species denoted 
by $n_i$ with $i=1,\ldots,N/2$ with two members in 
each species~(see Fig.~\ref{fig1}).
For each node $x$, a cluster $\mathcal{C}_x$ is defined as a
set of nodes that can be reached from $x$ via links.
We also define a $r$-neighbor $\mathcal{V}_x$ 
as a subset of $\mathcal{C}_x$ consisting of at most $[N^r]$ nodes. 
Here $0\leq r\leq 1$ and $[N^r]$ denotes the integer part of $N^r$.
It includes nodes in the ascending order of the distance from $x$.
If the size of $\mathcal{C}_x$ is less than $[N^r]$, then
$\mathcal{V}_x$ is simply given by $\mathcal{C}_x$. So, the sizes of
$\mathcal{V}_x$ and $\mathcal{C}_x$ denoted by $|\mathcal{V}_x|$ and
$|\mathcal{C}_x|$, respectively, satisfy
\begin{equation}\label{|V|}
|\mathcal{V}_x| = \min\{ | \mathcal{C}_x | , [N^r] \} \ .
\end{equation}

Initially, one starts with $N$ isolated nodes with no link. 
Then, links are added sequentially in the following way:
One selects two nodes $x$ and $y$ at random.
Two nodes are connected with a link if their $r$-neighbors
$\mathcal{V}_x$ and $\mathcal{V}_y$ do not share any
exclusive pair. Otherwise, linking is rejected~(see Fig.~\ref{fig1}).
The parameter $r$ represents a strength of the nonlocal constraint.

We are interested in how the onset of a percolation 
and the scaling of a percolating cluster
are affected by the nonlocal constraint parameterized by $r$.
The relevant measure is the sample-averaged largest-cluster size
$G$ as a function of the number of added links $L$ or the link density
$l=L/N$.  The percolation order parameter is given by $g \equiv G/N$. 
The mean cluster size $S$ is also measured~\cite{Stauffer94}.

\begin{figure}
\includegraphics*[width=\columnwidth]{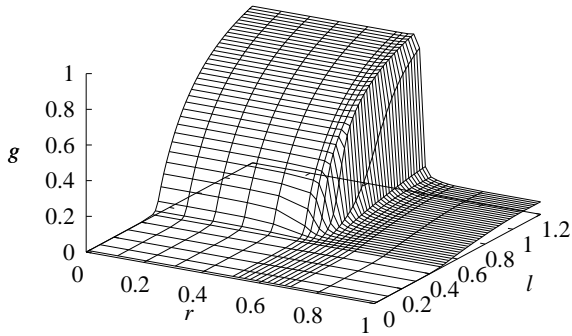}
\caption{Order parameter of the system with $N=64000$.}
\label{fig2}
\end{figure}

Figure~\ref{fig2} shows the overall behavior of the order parameter measured
in systems with $N=64 000$ nodes and averaged over $10000$
samples.
The order parameter displays a sharp transition
at small values of $r$, while it displays a weak transition behavior 
at large values of $r$.
Apparently the nonlocal constraint suppresses the emergence of a
percolating cluster. Percolation properties in both cases will be
studied in detail.

When $r=0$, the PE constraint becomes purely local in that
a linking trial is rejected only when selected nodes are an exclusive pair.
There are $N/2$ exclusive pairs among $N(N-1)/2$
possible selections. So a linking is rejected with the probability
\begin{equation}\label{p0}
p_0 = 1/(N-1) \ ,
\end{equation}
which can be ignored in the $N\to\infty$ limit.
Hence, the system becomes equivalent to the ER random network model in the
$N\to\infty$ limit, and undergoes a percolation transition in the MF
universality class at $l=l_c=1/2$.

With nonzero $r$, a linking probability $P_{x,y}$ for randomly selected
nodes $x$ and $y$ depends on
the size of their $r$-neighbors $\mathcal{V}_x$ and $\mathcal{V}_y$.
There are $|\mathcal{V}_x| \cdot |\mathcal{V}_y|$ possible combinations 
between nodes
and a linking trial is accepted only when none of them
are an exclusive pair. Hence, the linking probability is given by
\begin{equation}
P_{x,y} = (1-p_0)^{|\mathcal{V}_x|\cdot  | \mathcal{V}_y|} 
\end{equation}
with $p_0$ in Eq.~(\ref{p0}).
It is approximated as
\begin{equation}\label{Pxy}
P_{x,y} \simeq e^{-|\mathcal{V}_x| \cdot | \mathcal{V}_y| / N}
\end{equation}
in the large $N$ limit.

Equation~(\ref{Pxy}) gives a hint on the role of the nonlocal constraint.
We note that $|\mathcal{V}_x| \leq |C_x|$ from Eq.~(\ref{|V|}). Hence,
nodes belonging to a larger cluster have a higher rejection probability.
Consequently, the nonlocal constraint suppresses the growth of large
clusters. We also note that $|\mathcal{V}_x|\leq [N^r]$ from
Eq.~(\ref{|V|}). So, the quantity in the exponent of Eq.~(\ref{Pxy})
is bounded as
\begin{equation}\label{bound}
|\mathcal{V}_x| \cdot |\mathcal{V}_y| / N \leq N^{2r-1}
\end{equation}
for any pair of $x$ and $y$.
It indicates that the nonlocal constraint leads to a
different effect depending on whether $r<1/2$ or not.

Firstly, we consider the case with $r<1/2$. It will be referred to as
a {\em weak constraint} region.
In this region, the linking probability in Eq.~(\ref{Pxy})
converges to $1$ with a small correction 
of the order of $\mathcal{O}(N^{-(1-2r)})$ at most. Therefore,
the nonlocal constraint is negligible and the model is expected to display
the percolation transition belonging
to the MF universality class. Furthermore, 
the percolation threshold is equal to that of the ER random networks, 
$l_c(r)=1/2$.

\begin{figure}
\includegraphics*[width=\columnwidth]{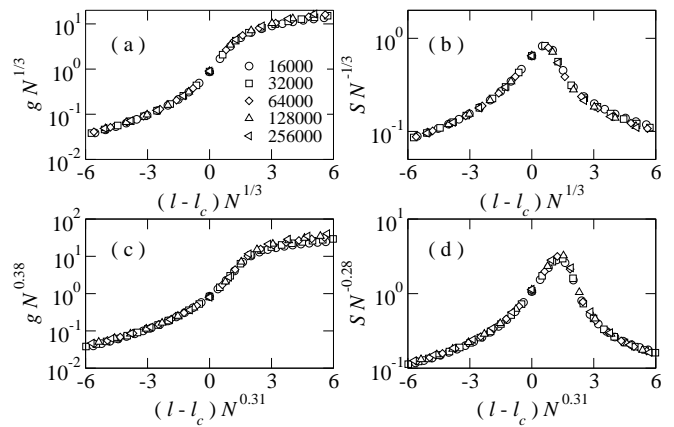}
\caption{FSS of the order parameter $g$ and the mean
cluster size $S$ at $r=0.3$ in (a) and (b) and at $r=0.5$
in (c) and (d). }
\label{fig3}
\end{figure}

In order to characterize the percolation transition, we have measured
numerically the order parameter $g$, the density of the largest cluster, and
the mean cluster size $S$. They satisfy the finite-size-scaling~(FSS)
forms
\begin{eqnarray}
g(l,N) &=& N^{-\beta/\bar{\nu}} \mathcal{F}_g ( (l-l_c) N^{1/\bar{\nu}})
\label{FSS_g} \ , \\
S(l,N) &=& N^{\gamma/\bar{\nu}} \mathcal{F}_S ( (l-l_c) N^{1/\bar{\nu}}) 
\label{FSS_S}
\end{eqnarray}
with the order parameter exponent $\beta$, the mean cluster size exponent
$\gamma$, and the FSS exponent $\bar{\nu}$~\cite{Stauffer94}. 
The MF universality class is characterized with the exponents
$\beta_{MF} = 1$, $\gamma_{MF}=1$, and
$\bar\nu_{MF}=3$~\cite{Lee04}.
Figures~\ref{fig3}(a) and (b) present the scaling plots for $g$ and $S$
at $r=0.3$.
Each data set collapses well onto a single curve with
$l_c=1/2$ and the MF critical exponents.
We obtained the same scaling behavior for other values of $r<1/2$.
This confirms our expectation based on the scaling of the linking
probability.

Secondly, we consider the case with $r>1/2$, which will be referred to as a
{\em strong constraint} region.
Here, the scaling behavior of $P_{x,y}$ in Eq.~(\ref{Pxy})
depends on the cluster size distribution. Since the PE constraint suppresses
the growth of the largest cluster, its size $G$ is bounded above by
the largest cluster size of the ER random network. To a given value of 
$l<1/2$, the ER random network displays a subcritical scaling 
$G \sim \ln N$~\cite{Bray88}. Consequently, the linking probability
$P_{x,y}$ for any $x$ and $y$ converges to $1$ with a correction 
of the order of $\mathcal{O}((\ln N)^2 / N)$ at most. 
Therefore, for $l<1/2$, the system becomes equivalent to the ER random 
network model and lies in the non-percolating phase with
$g \sim \ln N/ N$.

In the strong constraint case, the constraint is irrelevant 
for $l<1/2$ since $|\mathcal{V}_x| \leq |\mathcal{C}_x| \leq  G  = 
\mathcal{\ln N}$ for all $x$.
It becomes relevant when the largest cluster size
reaches a scaling $G\sim N^{1/2}$.
One can locate the threshold value $l_c(r)$ at which $G$ follows a critical
power-law scaling from the following scaling argument:
The subcritical scaling $G\sim \ln N$ for $l<1/2$ in the previous paragraph
implies that $l_c(r)\geq 1/2$.
In the ER random network model~($r=0$), the largest cluster
follows the critical scaling $G\sim N^{1-\beta_{MF}/\bar\nu_{MF}} = N^{2/3}$
at its percolation threshold $l=1/2$~\cite{Lee04}. The exponent value is
larger than $1/2$. This implies that the nonlocal constraint is already 
relevant at $l=1/2$ suggesting that $l_c(r) \leq 1/2$.
Therefore, the threshold value is given by $l_c(r) = 1/2$ at all values of
$r$. Furthermore, we expect that the critical largest cluster size
follows a power-law scaling
\begin{equation}
G_c(N) \sim N^{\alpha_c}
\end{equation}
with $1/2 \leq \alpha_c \leq 2/3$ in the strong constraint case. 

\begin{figure}
\includegraphics*[width=\columnwidth]{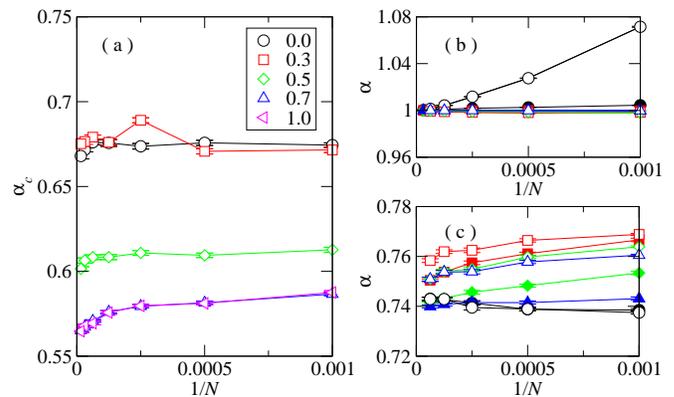}
\caption{(Color online) (a) Effective exponent $\alpha_c(N)$ for the critical
power-law scaling of $G_c \sim N^{\alpha_c}$ at the percolation threshold
$l=1/2$ at several values of $r$ as indicated by legends.
(b) Effective exponent $\alpha(N)$ for the power-law scaling of $G\sim
N^\alpha$ at $r=0.3$~(filled symbols) and $r=0.5$~(open symbols).
(c) Effective exponent $\alpha(N)$ for the power-law scaling of $G\sim
N^\alpha$ at $r=0.7$~(filled symbols) and $r=1.0$~(open symbols).
In (b) and (c), link densities are $l=0.75$~($\circ$), $l=1.0$~($\Box$),
$l=1.25$~($\diamond$), and $l=2.0$~($\triangle$).
}
\label{fig4}
\end{figure}

We have measured numerically the largest cluster size $G$ at $l=l_c$
to estimate the exponent $\alpha_c$.
Figure~\ref{fig4}(a) shows the numerical data for
the effective exponent $\alpha_c(N) \equiv \ln( G_c(2N) / G_c(N)) / \ln 2$
at several values of $r$. We find that
\begin{equation}
\alpha_c = 0.56 (5)
\end{equation}
for $r>1/2$. The exponent value is different from the MF value
$(1-\beta_{MF}/\bar{\nu}_{MF}) = 2/3$. 
Note that $\alpha_c \simeq 2/3$ in the weak constraint region.
The marginal case with $r=1/2$ will be considered later.

Can the system have a macroscopic percolating cluster of size 
$G = \mathcal{O}(N)$ in the 
strong constraint case~($r>1/2$)? Let us recall the mechanism
leading to a percolating cluster in the random network model~($r=0$).
As one adds links between pairs of nodes, clusters
merge with each other and grow. Nodes are selected for linking at random
with the same probability. However, growth rates of clusters are not
uniform. Larger clusters grow faster than smaller ones  because a cluster
is chosen for linking with the probability proportional to its size.
Hence, once there emerges a dominant cluster whose size is considerably
larger than the others, it grows even faster and eventually forms a
giant cluster.

The strong constraint suppresses the growth of large clusters severely. 
Especially, the linking probability between clusters of size
$\mathcal{O}(N^{q})$ with $q>1/2$ is vanishingly small. 
It suggests that the strong constraint does
not allow for a single dominant cluster. Then, a possible scenario is
that there appears a set of mesoscopic clusters of a same FSS behavior.
It is reminiscent of a so-called powder keg
scenario for the explosive percolation~\cite{Friedman09}.
Explosive percolation models also have a nonlocal
constraint suppressing the growth of large clusters. It results in
a powder keg of large clusters, which suddenly merge into a giant cluster
at a percolation threshold. The constraint in our model is much stronger.
For example, once a linking between two nodes is rejected, they are never
connected afterwards. It suggests that a giant cluster of size
$G = \mathcal{O}(N)$ may be improbable. Detailed properties are
investigated numerically.

The size of the largest cluster $G$ at $l>l_c$ is measured numerically.
It turns out to follow a power-law scaling
\begin{equation}\label{G_N_alpha}
G \sim N^\alpha \ .
\end{equation}
Numerical values of the exponent are estimated by using an
effective exponent $\alpha(N) \equiv \ln[G(2N)/G(N)]/\ln 2$.
Figures~\ref{fig4}(b) shows the plot of $\alpha(N)$ at $r=0.3$. 
It converges to $1$ for all values of $l>l_c$ indicative of a macroscopic
percolating cluster.
On the other hand, Fig.~\ref{fig4}(c) shows that the exponent converges to
\begin{equation}
\alpha = 0.75 (1)
\end{equation}
at all values of $r>1/2$. Therefore we conclude that the strong
constraint leads to a quasi-critical phase characterized with the exponent
$\alpha \simeq 0.75$. 

In the marginal case with $r=1/2$, the system displays a distinct
percolation transition. The largest cluster
size scales as $G \sim N^{\alpha_c}$ with $\alpha_c
= 0.61(2)$~(see Fig.~\ref{fig4}(a)) at the percolation threshold and 
there exists a macroscopic giant cluster with $G \sim N^1$ 
in the supercritical phase~(see Fig.~\ref{fig4}(b)). 
Using the FSS analysis based on Eqs.~(\ref{FSS_g}) and (\ref{FSS_S}), we
find that the transition is characterized 
with the non-MF 
critical exponents~(see Figs.~\ref{fig3}(c) and (d))
\begin{equation}\label{nonMF}
\beta/\bar\nu = 0.38(2) ,\ \gamma/\bar\nu = 0.28(3),\ 
1/\bar\nu = 0.31(3) \ .
\end{equation}
They satisfy the scaling relation $2\beta/\bar\nu +
\gamma/\bar\nu = 1$~\cite{Stauffer94} within error bars. 
These non-MF exponents indicates that the
marginal case constitutes a distinct universality class for percolation 
transitions.

\begin{figure}[t]
\includegraphics*[width=0.8\columnwidth]{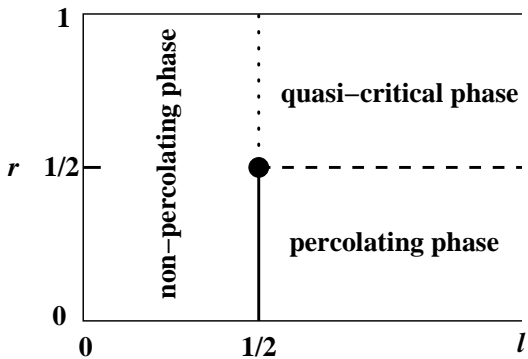}
\caption{Phase diagram}
\label{fig5}
\end{figure}

In summary, we have introduced a percolation model with a nonlocal constraint
and investigated the nature of percolation transitions numerically.
Our results are summarized in the phase diagram in
Fig.~\ref{fig5}. When $l<l_c = 1/2$, the system is always 
in the non-percolating phase
where $G \sim  \ln N$. When $l>l_c$, the system
is in the percolating phase with $G \sim N^1$ for $r\leq 1/2$
while it is in the quasi-critical phase with
$G \sim N^{\alpha}$ with $\alpha \simeq 0.75$
for $r>1/2$. The critical line $l = l_c$ in the weak constraint
region~(solid line) belongs to the MF universality class. The critical line
$l=l_c$ in the strong constraint region~(dotted line) separates the
non-percolating phase and the quasi-critical phase. The marginal case at
$r=1/2$ belongs to a distinct universality class characterized with the
non-MF exponents in Eq.~(\ref{nonMF}).
Our study shows that the nonlocal constraint leads to rich percolation
critical phenomena. It calls for further study of percolation
transitions in systems with nonlocal constraints.

\begin{acknowledgments}
This work was supported by the 2011 sabbatical year research grant of the
University of Seoul.
\end{acknowledgments}

\end{document}